\documentclass[11pt]{article}
\usepackage{graphicx}
\usepackage{epstopdf}
\usepackage{amsmath}
\usepackage{amssymb}
\usepackage{amsfonts}
\usepackage{amsthm}
\usepackage[usenames]{color}
\usepackage{array}

\usepackage[
      colorlinks=true,
      linkcolor=blue,
      urlcolor=blue,
      filecolor=blue,
      citecolor=red,
      pdfstartview=FitV,
      pdftitle={},
      pdfauthor={},
      pdfsubject={},
      pdfkeywords={},
      pdfpagemode=None,
      bookmarksopen=true
]{hyperref}

\usepackage{epsfig}

\usepackage{hyperref}

\def\d{\partial}

\def\beq{\begin{eqnarray}}
\def\eeq{\end{eqnarray}}

\def\a{\alpha}
\def\b{\beta}
\def\c{\gamma}
\def\d{\delta}
\def\e{\epsilon}
\def\f{\sigma}

\def\m{\mu}

\def\x{\xi}

\def\be{\begin{equation}}
\def\ee{\end{equation}}
\def\bea{\begin{eqnarray}}
\def\eea{\end{eqnarray}}

\def\be{\begin{equation}}
\def\ee{\end{equation}}
\def\bea{\begin{eqnarray}}
\def\eea{\end{eqnarray}}

\newcommand{\rom}[1]{\mathrm{#1}}

\def\cH{\mathcal{H}}

\def\nn{\nonumber}

\numberwithin{equation}{section}

\begin{document}

\begin{centering}

\textbf{\Large{On Propagation of Energy  Flux in de Sitter Spacetime}}

 \vspace{0.8cm}

Sk Jahanur Hoque$^1$ and Amitabh Virmani$^{1,2,3, }$\footnote{Currently on lien from Institute of Physics, Sachivalaya Marg, Bhubaneswar, Odisha, India 751005. }

  \vspace{0.5cm}

 \vspace{0.5cm}

\begin{minipage}{.9\textwidth}\small  \begin{center}
$^1$Chennai Mathematical Institute, H1 SIPCOT IT Park, \\ Kelambakkam, Tamil Nadu, India 603103\\
  \vspace{0.5cm}
$^2$Institute of Physics, Sachivalaya Marg, \\ Bhubaneswar, Odisha, India 751005 \\
  \vspace{0.5cm}
$^3$Homi Bhabha National Institute, Training School Complex, \\ Anushakti Nagar, Mumbai 400085, India \\
  \vspace{0.5cm}
{\tt skjhoque, avirmani@cmi.ac.in}
\\ $ \, $ \\

\end{center}
\end{minipage}

\end{centering}

\begin{abstract}
In this paper, we explore propagation of energy flux in the future Poincar\'e patch of de Sitter spacetime. We present two results. First, we compute the flux integral of energy using the symplectic current density of the covariant phase space approach 
on hypersurfaces of constant radial physical distance. Using this computation we show that in the tt-projection,  the integrand in the energy flux expression on the cosmological horizon is same as that on the future null infinity. This suggests that propagation  of energy flux in de Sitter spacetime is sharp. Second, we relate our energy flux expression in tt-projection to a previously obtained expression using the Isaacson stress-tensor approach. 
\end{abstract}

\newpage

\tableofcontents

\section{Introduction}

The era of gravitational wave astronomy  has begun
\cite{Abbott:2016blz, TheLIGOScientific:2017qsa, GBM:2017lvd, Coulter}. It is now all the more important that our theoretical understanding be at par with the impressive experimental developments that have gone into the discovery of gravitational waves. There are several theoretical aspects that are potentially important in relation to generation and propagation of gravitational waves but have not been fully explored. One such aspect is the effect of the positive cosmological constant on the propagation of gravitational waves.

The discovery of the accelerated expansion of the universe from distant supernovae and cosmic microwave background surveys have shown that around 68\% of the energy density of the universe is dark energy. While  at a fundamental level dark energy is poorly understood, the  positive cosmological constant is the simplest explanation of it. From the theoretical point of view, positive cosmological constant posses numerous challenges in relation to study of gravitational waves. In a recent series of papers Ashtekar, Bonga, and Kesavan \cite{ABKI, ABKII, ABKIIIPRL, ABKIII} have systematically initiated the study of gravitational waves focusing on the numerous effects that the presence of a positive cosmological constant brings.  Subsequently, several authors have contributed to the development of the subject \cite{DHI, DHII, Bishop:2015kay, Bonga, JA}.   The primary aim of this work is to expand on some of these studies, in particular on some aspects of \cite{DHI, DHII}, and to clarify their relation to \cite{ABKII, ABKIII}.

 In comparison to Minkowski spacetime there are several effects that the positive cosmological constant brings on the propagation of linearised gravitational field. For a detailed discussion of these points, we refer the reader to \cite{ABKII, ABKIII}; here we wish to focus on two points especially. First, while wavelengths of linear waves remain constant in flat space, they increase in de Sitter spacetime as the universe undergoes de Sitter expansion. So much so that in the asymptotic region of interest, the wavelengths diverge. Naively, this seems to invalidate the geometrical optics approximation commonly used in the gravitational waves literature. Secondly, due to the curvature of the background spacetime, the linear gravitational field satisfies a  \emph{massive} wave equation, i.e., propagation of waves in de Sitter spacetime is not on the light cone. Due to backscattering from the background curvature, in general, there is a tail term.

Partial understanding of  these effects is already available. Our study expands on that knowledge. Firstly, 
although in the asymptotic region of interest, wavelengths diverge, reference \cite{DHII}  made precise how the geometrical optics approximation is still useful.
They arrived 
at an effective stress tensor  for gravitational waves following the original work of Isaacson \cite{Isaacson:1968zza, Isaacson:1967zz}. 
An aim of this paper is to re-obtain appropriate version of those expressions from the covariant phase space approach, thus clarifying their relation to \cite{ABKII, ABKIII}.
 The second aim of the paper is to make precise the notion of  the ``sharp" propagation of energy flux in de Sitter spacetime, i.e., to understand in what sense the tail term mentioned above does not matter for radiated energy flux.

The rest of the paper is organized as follows. We start with a brief review of linearised gravity on de Sitter spacetime in section \ref{sec:linear} and write various identities involving derivatives of the radiative field that we need in later sections. In section 
\ref{sec:current_and_energy} we compute the symplectic current  density for linearised gravity on de Sitter spacetime and write a general expression for the energy flux through a hypersurface $\Sigma$. Since symplectic current  density is conserved,  it allows us to compute energy flux through any hypersurface.

In section \ref{sec:tt} we use the general expression obtained in section
\ref{sec:current_and_energy} to compute the flux integrals on hypersurfaces of constant radial physical distance. These hypersurfaces allow us to interpolate between the cosmological horizon and the future null infinity. We show that in the tt-projection,  the integrand in the energy flux expression on the cosmological horizon is same as that on the future null infinity. This suggests that the propagation of energy flux in de Sitter spacetime is sharp.
We also relate our energy flux expression to the previously obtained expression of reference \cite{DHII}.
This section constitutes the main results of our work. 

We close with a discussion in section \ref{sec:disc}.

 \begin{figure}[t]
\centering
\begin{center}
\includegraphics[width=0.7\textwidth]{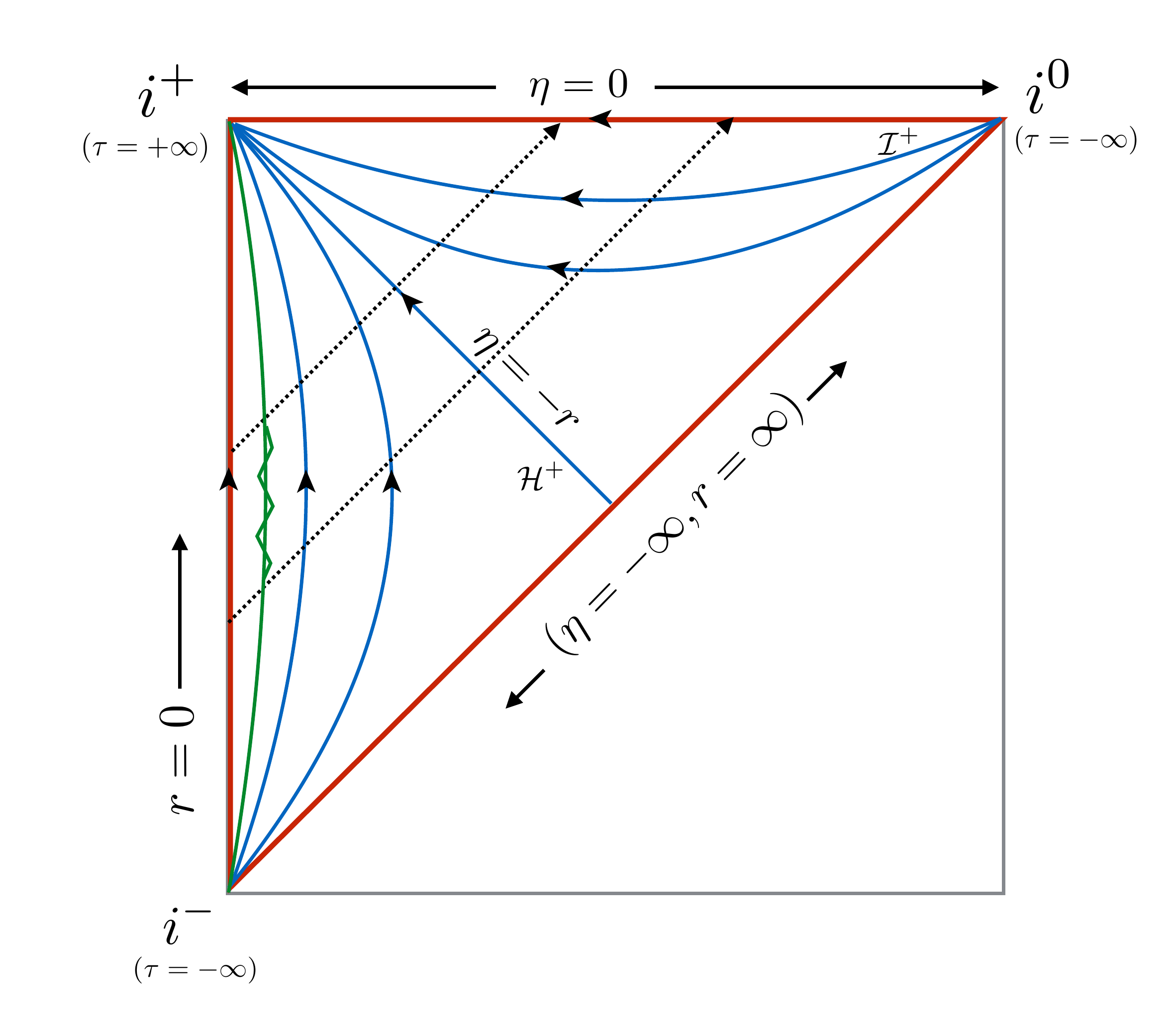}
\caption{The full square is the Penrose diagram of global de Sitter spacetime, 
with each point representing a 2-sphere. In this paper we exclusively work in the future Poincar\'e patch  of de Sitter spacetime --- the upper triangular region (red triangle) of this diagram.
Blue lines denote hypersurfaces of constant radial physical distance. 
These hypersurfaces are generated  by the time-translational (dilatation) Killing vector. On these hypersurfaces, $\tau$ is a Killing parameter that runs from $-\infty$ to $\infty$. The dotted lines are lines of constant retarded time. Green line is the worldline of the radiating source. }
\label{PoincarePatch}
\end{center}
\end{figure}

\section{Linearised gravity and various identities involving radiative field}

\label{sec:linear}
\subsection{Linearised gravity on de Sitter spacetime}
We are interested in linearised gravity over de Sitter background. We exclusively work in the future Poincar\'e patch of de Sitter spacetime.  The background de Sitter metric in the Poincar\'e patch is 
\be
ds^2 = \bar g_{\a \b} dx^\a dx^\b = a^{2} ( - d\eta^2 + d\vec x^2),  \label{background}
\ee
\be
a=-(H\eta)^{-1}, \quad \mbox{with} \quad H=\sqrt{\Lambda/3},
\ee
where $\Lambda$ is the positive cosmological constant. 
Linearised perturbations over the background \eqref{background} are written as
\be
g_{\a \b} = \bar g_{\a \b} +  \gamma_{\a \b}.
\ee 
Coordinates $x_i$, with $i = 1,2,3$, ranges from $(-\infty, \infty)$, whereas coordinate $
\eta$ takes values in the range $(-\infty, 0)$, with $\eta = 0$ at the future 
null infinity $\mathcal{I}^+$. The future 
infinity $\mathcal{I}^+$ is a spacelike surface, see figure~\ref{PoincarePatch}.

For the background metric the Christoffel symbol is 
\be
\overline{\Gamma}^\a_{\b \c} = - \frac{1}{\eta} \left( \delta^0_\b \delta^\a_\c + \delta^0_\c \delta^\a_\b + \delta^\a_0 \eta_{\b \c}\right).
\ee
Using this useful expressions for the d'Alembertian and for various other derivative operators can be written,  see e.g. \cite{deVega:1998ia, DHI}.  
In terms of the trace reversed combination 
$\hat{\gamma}_{\alpha\beta}:=\gamma_{\alpha\beta}-\frac{1}{2}\bar{g}
_{\alpha\beta} \ (\bar{g}^{\mu\nu}\gamma_{\mu\nu})$, the linearised 
Einstein equations take the form,
\begin{equation} 
\frac{1}{2}\left[ - \overline{\Box} \hat{\gamma}_{\mu\nu} + \left\{
\overline{\nabla}_{\mu}B_{\nu} + \overline{\nabla}_{\nu}B_{\mu} -
\bar{g}_{\mu\nu}(\overline{\nabla}^{\alpha}B_{\alpha})\right\}\right] +
\frac{\Lambda}{3}\left[\hat{\gamma}_{\mu\nu} -
\hat{\gamma}\bar{g}_{\mu\nu}\right] ~ = ~ 8\pi G T_{\mu\nu} \label{LinEqn}
\end{equation}
where $B_{\mu} := \overline{\nabla}_{\alpha}\hat{\gamma}^{\alpha}_{~\mu}$ and   $\overline{\nabla}_{\alpha}$ is the metric compatible covariant derivative with respect to the background metric $\bar g_{\a \b}$.

As is well known in the literature \cite{deVega:1998ia, DHI, ABKIII}, these equations written in terms of a rescaled variables leads to a 
 great deal of 
simplification. We define, 
\be
\chi_{\mu
\nu}:=a^{-2}\hat{\gamma}_{\mu\nu},
\ee 
and using the gauge condition \cite{deVega:1998ia},
\be
\partial^{\alpha}\chi_{\alpha\mu} + \frac{1}{\eta}\left(2
\chi_{0\mu} + \delta_{\mu}^0 \chi_{\alpha}^{~\alpha} \right) = 0,
\label{ChiGauge}
\ee
equation \eqref{LinEqn} becomes,
\be
-~16 \pi G T_{\mu\nu} = \Box \chi_{\mu\nu} +
\frac{2}{\eta}\partial_0\chi_{\mu\nu} -
\frac{2}{\eta^2}\left(\delta_{\mu}^0\delta_{\nu}^0\chi_{\alpha}^{~\alpha}
+ \delta_{\mu}^0\chi_{0\nu} + \delta_{\nu}^0\chi_{0\mu} \right)
\label{LinEqnChi},
\ee
where $\Box$ is simply the d'Alembertian with respect to Minkowski metric in cartesian coordinates, $\Box = - \partial_\eta^2 + \partial_i^2$.

In terms of variables
$
\hat{\chi} :=\chi_{00}+\chi_{~i}^{i}, ~\chi_{0i},~ \chi_{ij}
$
equation (\ref{LinEqnChi}) decomposes into three decoupled equations
\bea
\Box \bigg(\frac{\hat{\chi}}{\eta}\bigg) &=& - \frac{16 \pi G \hat{T}}{\eta}, \label{decoupled1}\\
\Box \bigg(\frac{{\chi_{0i}}}{\eta}\bigg)&=&- \frac{16 \pi G {T}_{0i}}{\eta}, \label{decoupled2} \\
\Box \chi_{ij} + \frac{2}{\eta}\partial_0  \chi_{ij} &=& - 16\pi G T_{ij}, \label{decoupled3}
\eea
where $\hat{T}:=T_{00}+T_{~i}^{i}$.

Under a linearised diffeomorphism  $\xi^{\mu}$, $\chi_{\mu\nu}$ transforms as, 
\be
\delta\chi_{\mu\nu} = (\partial_{\mu}\underline{\xi}_{\nu} +
\partial_{\nu}\underline{\xi}_{\mu} -
\eta_{\mu\nu}\partial^{\alpha}\underline{\xi}_{\alpha}) -
\frac{2}{\eta}\eta_{\mu\nu}\underline{\xi}_0,
\ee
where
\be
\underline{\xi}_{\mu}
:= a^{-2} \xi_{\mu} = \eta_{\mu\nu}\xi^{\nu}.  \label{ChiGaugeTrans}
\ee
 A small calculation then shows that the gauge condition (\ref{ChiGauge}) is preserved under transformations
generated by  vector fields $\xi^{\mu}$ --- the residual gauge transformations --- satisfying, 
\begin{equation}
\Box\underline{\xi}_{\mu} +
\frac{2}{\eta}\partial_0\underline{\xi}_{\mu} -
\frac{2}{\eta^2}\delta_{\mu}^0\underline{\xi}_0= 0.
\label{ResidualChiGauge}
\end{equation}
Under these residual gauge transformations
equation \eqref{LinEqnChi} 
is also invariant.

We can exhaust the residual gauge freedom as follows. We note that  $\delta{\hat{\chi}}$ satisfies,
\begin{eqnarray}
\Box~{\big(\delta{\hat{\chi}}\big)}&=& 4~\bigg[\Box~\bigg(\partial_0 \underline{\xi}_0-
\frac{\underline{\xi}_0}{\eta}\bigg)\bigg] \\
&=&-\frac{2}{\eta}\partial_0~
\big(\delta{\hat{\chi}}\big)+\frac{2}{\eta^{2}}~\big(\delta{\hat{\chi}}\big) 
\eea
where in going from the first step to the second step we have used \eqref{ResidualChiGauge}. This form of the equation implies that,
\be 
\Box~\bigg(\frac{\delta{\hat{\chi}}}{\eta}\bigg)=0,
\ee
i.e., $\delta{\hat{\chi}}$ satisfies the wave equation \eqref{decoupled1} outside the source. 
Therefore, using an appropriate residual gauge transformation we can set $
\hat{\chi}=0$ 
outside the source.  Similarly $\chi_{0i}$ can be set to zero outside the source \cite{deVega:1998ia}.

Gauge condition (\ref{ChiGauge}) then implies
$\partial^0\chi_{00} = 0$. Choosing $\chi_{00}$ to be zero at some
initial $\eta = $ constant hypersurface we can take $\chi_{00} = 0$ everywhere. Doing so, 
gauge condition (\ref{ChiGauge})  becomes 
\be
\partial^{i}\chi_{ij} = \chi^{i}_{~i} = 0. \label{TTconditions}
\ee

\subsection{TT-gauge vs tt-projection}

With conditions \eqref{TTconditions} imposed there are no further gauge transformations allowed. 
Thus,  transverse and 
traceless (TT) solutions are fully gauge fixed.  Therefore, away from the source it suffices to focus on equation (\ref{decoupled3}). 
In general, solutions of this inhomogeneous equation do not satisfy the TT conditions. However, any 
spatial rank-2 symmetric tensor can be decomposed into its irreducible components as,
\begin{equation}
\chi_{ij}=\frac{1}{3}\delta_{ij}\delta^{kl}\chi_{kl}+(\partial_{i}\partial_{j}-
\frac{1}{3}\delta_{ij}
\nabla^{2})B+\partial_{i}B_{j}^\rom{T}+\partial_{j}B_{i}^\rom{T}+\chi_{ij}^\rom{TT},
\end{equation}
where $\chi_{ij}^\rom{TT} $ refers to the transverse-traceless part of the field $\chi_{ij}$, i.e., it satisfies,  
\be
\partial^{i}
\chi_{ij}^\rom{TT}=\delta^{ij} \chi_{ij}^\rom{TT}=0.
\ee 
The vector $B_{i}^\rom{T}$ is transverse,
$\partial^i B_i^\rom{T} = 0.$ In this decomposition only $
\chi_{ij}^\rom{TT}$ is
the gauge invariant piece. Hence, $\chi_{ij}^\rom{TT}$ is best regarded as the physical 
component of the field $\chi_{ij}$. 
Given a tensor $\chi_{ij}$, in general it is highly non-trivial to extract $\chi_{ij}^\rom{TT}$; see \cite{Bonga} for an explicit example. 

In the context of gravitational waves, another conceptually distinct notion of transverse-traceless tensors is often used in the literature. This notion is operationally simpler but inequivalent to the above notion.  Here one `extracts' the `transverse-traceless' part of a rank-two tensor simply  by defining an algebraic projection operator, 
\begin{align}
P_i^{~j}  &=  \delta_i^{~j} - \hat{x}_i\hat{x}^{j}, &
\Lambda_{ij}^{~~kl}  &=  \frac{1}{2}(P_i^{~k}P_j^{~l} +
P_i^{~l}P_j^{~k} - P_{ij}P^{kl}),
\end{align}
where $\hat x^i = x^i/r$ with $r= \sqrt{x^i x_i}$.
In order to distinguish it from the the above notion, we use the notation $\chi_{ij}^\rom{tt}$, 
\be
\chi_{ij}^\rom{tt}  := \Lambda_{ij}^{~~kl}\chi_{kl} 
\label{ttProjection}
\ee
For a detailed discussion of the differences between these two notions see \cite{Ashtekar:2017ydh, Ashtekar:2017wgq}. For asymptotically flat 
space-times the two notions match only at null infinity $\mathcal{I}^+$ \cite{Ashtekar:2017wgq, Bonga}. The tt-projection is well tailored to the $1/r$ expansion commonly used for asymptotically flat spacetimes.

The global structure of  de Sitter spacetime is very different from Minkowski spacetime. Expansion in powers of $1/r$ is not a useful tool to analyse asymptotically de Sitter spacetimes. In particular, the radial tt-projection is not a valid operation to extract the transverse-traceless part of a rank-2 tensor on the  full $\mathcal{I}^+$. The TT-tensor is the correct notion of transverse traceless tensors. However, if one restricts oneself to large radial distances away from the source,  one may expect that the tt-projection also gives useful answers. In fact, it appears to work better than expected. In 
the context of the power radiated by a spatially compact circular binary system analysed in 
\cite{Bonga,JA} the difference does not seem to matter. 

The tt-projection being algebraic allows us to do various non-trivial computations which seem difficult to perform otherwise. In particular, this simplicity allows us to gain a physical understanding of the propagation of gravitational waves in de Sitter spacetime. In this paper we mostly restrict ourselves to tt-projection, with the understanding that our results need to be generalised to TT gauge.  A detailed study of this we leave for future research. %some preliminary comments are made in the appendix. 

%%%%%%%%%%%%%%%%%%%%%%%%%%%%%%%%%%%%

\subsection{Derivatives of radiative field}

\label{Identity}

In order to compute energy flux through different slices, we need  various 
derivatives of radiative field $\chi_{ij}$. In this subsection we establish those identities. 
The expression for radiative $\chi_{ij}$ we use was obtained in references \cite{ABKIII, DHI},
\be
\chi_{ij} (\eta, r) = 4 G\frac{\eta}{r (\eta - r)} \int d^3 x' T_{ij}(\eta - r, x') + 4 G \int_{-\infty}^{\eta - r} d\eta' \frac{1}{\eta'{}^2}\int d^3 x' T_{ij}(\eta', x'),
\ee
where $T_{ij}$ is the source energy-momentum tensor. In arriving at this expression, Green's functions for the differential operator in  \eqref{decoupled3} is used together with the approximation
\be
\eta - | \vec x - \vec x'| \approx \eta - | \vec x| = \eta - r,
\ee
in order to pull the factor of $\frac{1}{r(\eta - r)}$ out from the integral. 
The integral of the stress tensor can be expressed in terms of the mass and pressure quadrupole moments $Q_{ij}$ and $\overline{Q}
_{ij}$ at the retarded time $\eta_\rom{ret}:=\eta - r$ \cite{ABKIII, DHI},
\be
\int d^3 x' T_{ij}(\eta - r, x') = \frac{1}{2 a(\eta_\rom{ret})} \left( \ddot{Q}_{ij} + 
2 H \dot Q_{ij} + 2 H \dot{\overline{Q}}_{ij} + 2 H^2 {\overline{Q}}
_{ij} \right) (\eta_\rom{ret}), \label{moments}
\ee
where dots denote Lie derivatives with respect to time-translation (dilatation) Killing vector 
\be
T^{\mu} \partial_\mu = - H ( \eta \partial_\eta + r \partial_r). \label{time_translation}
\ee

The mass and pressure quadrupole moments $Q_{ij}$ and $\overline{Q}
_{ij}$ are defined as an integrals over the source at some fixed time $\eta$,
\bea
Q_{ij} (\eta)= \int \ a^3 (\eta) T_{00} (\eta, x) x_i  x_j d^3 x,
\eea
\bea
\overline{Q}_{ij} (\eta)= \int \ a^3 (\eta) \delta^{kl}T_{kl}(\eta, x) x_i  x_j d^3 x.
\eea

Using these expressions, we get the identities
\be
\partial_\eta \chi_{ij} (\eta, x) = 4G\frac{\eta}{(\eta - r)r} \partial_\eta 
\left[ \int d^3 x' T_{ij}(\eta - r, x') \right] =: \frac{2G\eta}{(\eta - r)r} 
R_{ij} (\eta_\rom{ret}).
\ee
where
\be
R_{ij}(\eta_\rom{ret}) ~ = ~\bigg[\dddot{Q}_{ij} + 3H\ddot{Q}_{ij} +
2H^2\dot{Q}_{ij} + H\ddot{\overline{Q}}_{ij} + 3H^2\dot{\overline{Q}}_{ij} +
2H^3\overline{Q}_{ij}\bigg](\eta_\rom{ret}) .
\ee
Similarly, 
\bea
\partial_r \chi_{ij} &=& - \partial_\eta \chi_{ij} - \frac{4}{r^2}\int d^3 x' T_{ij}
(\eta - r, x') \label{useful_identity_DH}
\\
&=&-\frac{2G\eta}{r (\eta - r)}  R_{ij}(\eta_\rom{ret})+ 2H G \ \frac{(\eta -r)}{r^{2}}\left( 
\ddot{Q}_{ij} + 2 H \dot Q_{ij} + 2 H \dot{\overline{Q}}_{ij} + 2 H^2 
\overline{Q}
_{ij} \right)(\eta_\rom{ret}).
\eea
As a result
\bea
(T \cdot \partial) \chi_{ij} &=& - H (\eta \partial_\eta + r \partial_r ) \chi_{ij} \\
&=&-\frac{2G H \eta}{r} R_{ij} (\eta_\rom{ret}) -2GH^{2} \bigg(\frac{\eta - r }{r}\bigg)\left( 
\ddot{Q}_{ij} + 2 H \dot Q_{ij} + 2 H \dot{\overline{Q}}_{ij} + 2 H^2 
\overline{Q}_{ij} \right) (\eta_\rom{ret}). \label{T_dot_partial_chi}
\eea

For later convenience we also define 
\be
  A_{ij}=\ddot{Q}_{ij}+2H\dot{Q}_{ij}+H\dot{\overline{Q}}_{ij}+2H^{2}
  \overline{Q}_{ij}. \label{A_def}
  \ee 
 This quantity is interesting as it satisfies the relations  
\be
  R_{ij}=\dot{A}_{ij}+HA_{ij}= (T \cdot \partial) A_{ij}-HA_{ij},
\label{R_A_old}
   \ee  
which we will need later.

On the future cosmological horizon of the source defined by
\be
\cH^+: \qquad \eta  + r =0,
\ee 
equation \eqref{moments} simplifies to,
\be
\left[\int d^3 x' T_{ij}(\eta - r, x')\right]\Bigg{|}_{\cH^+} =  (H r)\left( \ddot{Q}_{ij} + 2 H \dot Q_{ij} + 2 H \dot{\overline{Q}}_{ij} + 2 H^2 \overline{Q}_{ij} \right)  (\eta_\rom{ret}),
\ee
and equation \eqref{T_dot_partial_chi} simplifies to,
\be
(T \cdot \partial) \chi_{ij} \Big{|}_{\cH^+} = 2G H R_{ij}  (\eta_\rom{ret}) + 4G H^2 \left( \ddot{Q}_{ij} + 2 H \dot Q_{ij} + 2 H \dot{\overline{Q}}_{ij} + 2 H^2 \overline{Q}_{ij} \right) (\eta_\rom{ret}).
\ee

\section{Symplectic current density and energy flux}
\label{sec:current_and_energy}
We are interested in computing energy flux through any Cauchy surface and more generally through other surfaces. Perhaps the most convenient way to do this is via the covariant phase space approach.  For linearised gravity, the covariant phase space can be taken to be simply the space of solutions $\gamma_{ab}$ of the linearised Einstein's equations together with appropriate gauge conditions \cite{ABKII}. A standard procedure  \cite{ABR, LeeWald} then gives a symplectic structure.

When restricted to cosmological slices, the symplectic structure was computed and used in \cite{ABKII, ABKIII}. In this work we are interested in other slices.  In our discussion below we focus on the  symplectic current density and its integrals, rather than on the careful construction of the phase space itself. The phase space construction is somewhat subtle \cite{ABKII} due to certain divergences as the future null infinity $\mathcal{I}^+$ is approached. Some of our intermediate expressions below are formally divergent as the future null infinity is approached, however, our final answers are all finite and have a well defined limit at $\mathcal{I}^+$.

We start with an expression of symplectic current of linearised Einstein gravity with a cosmological constant, which we can evaluate on different slices. A convenient form is \cite{Hollands:2005wt},
 \be
\omega^\a =\frac{1}{32\pi G} \ P^{ \a \b \c \d \e \f } 
\left( \delta_1 g_{ \b \c } \overline{\nabla}_{ \d } \delta_2 g_{\e \f } - \delta_2 g_{\b \c } \overline{\nabla}_{\d} \delta_1 g_{\e \f }\right), \label{omega_exp_1}
\ee
where
\be
P^{\a \b \c \d \e \f} = \bar g^{\a \e} \bar g^{\f \b} \bar g^{\c \d} - \frac{1}{2} \bar g^{\a \d} \bar g^{\b \e} \bar g^{\f \c} - \frac{1}{2} \bar g^{\a \b} \bar g^{\c \d} \bar g^{\e \f} - \frac{1}{2} \bar g^{\b \c} \bar g^{\a \e} \bar g^{\f \d} + \frac{1}{2}  \bar g^{\b \c} \bar g^{\a \d} \bar g^{\e \f}.
\ee
We use the notation 
\bea
\delta_1 g_{\a \b} &=& \gamma_{\a \b}, \\
\delta_2 g_{\a \b} &=& \widetilde \gamma_{\a \b},
\eea
where $\gamma_{\a \b}$ and $\widetilde \gamma_{\a \b}$ are fully gauge fixed physical solutions of the (homogeneous) linearised Einstein equations. We take them to satisfy 
Lorentz and radiation gauge, 
\be
\overline{\nabla}^\a {\gamma}_{\a\b} = 0, \qquad \gamma_{0\a} = 0, \qquad \bar{g}^{\a\b}  \gamma_{\a\b} = 0. \label{gauge_gamma}
\ee
These gauge conditions are the same as \eqref{TTconditions}.
Since $\gamma_{\a\b}$ and $\widetilde \gamma_{\a\b}$ are both traceless, the last three terms in $P^{\a \b \c \d \e \f}$ do not contribute to the symplectic current $\omega^\a$. We effectively have
\be
P^{\a \b \c \d \e \f} = \bar g^{\a \e} \bar g^{\f \b} \bar g^{\c \d} - \frac{1}{2} \bar g^{\a \d} \bar g^{\b \e} \bar g^{\f \c}. \label{P_simple}
\ee
Expanding out the covariant derivatives in \eqref{omega_exp_1} in terms of the Christoffel symbols we get a simplified expression,
\bea
\omega^\a 
&=& \frac{1}{32\pi G} \  P^{\a \b \c \d \e \f} \gamma_{\b \c} \left(\partial_\d \widetilde \gamma_{\e \f}  - \overline{\Gamma}^\m_{\d \e} \widetilde \gamma_{\m \f} - \overline{\Gamma}^\m_{\d \f} \widetilde \gamma_{\e \m} \right)  - (1 \leftrightarrow 2),
\eea
with $P^{\a \b \c \d \e \f}$ given in \eqref{P_simple}.

\subsubsection*{Time component}
Using the simplified expressions above, the time component of the symplectic current is
\be 
\omega^\eta = \frac{1}{64\pi G} (H^2 \eta^2) \left(\gamma^{\b \c} \partial_\eta \widetilde \gamma_{\b \c} -\widetilde \gamma^{\b \c} \partial_\eta  \gamma_{\b \c}  \right). 
\ee
We note that due to the gauge conditions \eqref{gauge_gamma}, $\gamma_{\a \b}$ has only spatial components. In terms of the rescaled field $\gamma_{ij} = a^2 \chi_{ij}$,
we have 
\be \label{SympCur1}
\omega^\eta = \frac{1}{64 \pi G} (H^2 \eta^2) \left(\chi^{ij} \partial_\eta \widetilde \chi_{ij} -\widetilde \chi_{ij} \partial_\eta  \chi_{ij}\right). 
\ee
This expression matches with the corresponding expression in reference \cite{ABKII}. In such expressions TT superscript on $\chi_{ij}$ is implicit. 
\subsubsection*{Space components}
A similar calculation gives
\be \label{SympCur2}
\omega^i = \frac{1}{32\pi G}a^{-2} \delta^{ij} \left\{ \chi^{lm} \partial_m \widetilde \chi_{jl} - \frac{1}{2} \chi^{lm} \partial_j \widetilde \chi_{lm} - \widetilde \chi^{lm} \partial_m  \chi_{jl} + \frac{1}{2} \widetilde \chi^{lm} \partial_j  \chi_{lm} \right\}.
\ee

\subsubsection*{Energy flux}
From general results on the covariant phase space approach \cite{ABR, LeeWald, ABKII}, 
it follows that  the energy flux (Hamiltonian for time-translation symmetry $T$) is given as
\be
E_T (\gamma) = -\int_{\Sigma} \omega^\alpha (\gamma, \pounds_T \gamma)  \left(n_{\alpha}  \sqrt{h_\Sigma}d^3 \x\right), \label{ET}
\ee
where $h_\Sigma$ is the determinant of the induced metric on the slice $\Sigma$  with coordinates $\xi^i$ and $n^\alpha$ is the future directed  normal  vector to the slice $\Sigma$. In this expression we have evaluated the symplectic current density with $\widetilde \gamma_{\a\b} = \pounds_T \gamma_{\a\b}$.  For use in equations \eqref{SympCur1} and \eqref{SympCur2}, we need to evaluate $\widetilde \chi_{ij} = a^{-2} \widetilde \gamma_{ij} = a^{-2} \pounds_T \gamma_{ij}$. This quantity is computed to be
\bea
\widetilde \chi_{ij}  
= a^{-2}  \  \pounds_T (a^2 \chi_{ij}) 
= (T \cdot \partial) \chi_{ij}.
\eea
As a result we have the following components of the current $j^\a := \omega^\alpha (\gamma, \pounds_T \gamma)$ for computing the energy flux,
\bea
j^\eta &=&  \frac{1}{64\pi G} \ (H^2 \eta^2) \left(\chi^{ij} \partial_\eta \left[ (T \cdot \partial) \chi_{ij} \right]  - (T \cdot \partial) \chi_{ij}  \partial_\eta  \chi^{ij}\right), \label{EFI} \\
j^i &=& \frac{1}{32\pi G} \ (H^2 \eta^2) \delta^{ik} \left\{ \chi^{lm} \partial_m \left[ (T \cdot \partial) \chi_{kl} \right] - \frac{1}{2} \chi^{lm} \partial_k \left[ (T \cdot \partial) \chi_{lm} \right] \right. \nn  \\ 
& &\qquad \qquad  \qquad \qquad \left.  - \left[ (T \cdot \partial) \chi^{lm} \right] \partial_m  \chi_{kl} + \frac{1}{2} \left[ (T \cdot \partial) \chi^{lm} \right] \partial_k  \chi_{lm} \right\}.\label{EFII}
\eea
Since $j^\a$ is  conserved, we can use it to compute flux across any hypersurface. In this paper we will restrict ourselves to hypersurfaces generated by the time-translation Killing vector $T$. Near the future null infinity $\mathcal{I}^+$ these hypersurfaces are spacelike. Inside the cosmological horizon $\mathcal{H}^+$ these hypersurfaces are timelike. See figure~\ref{PoincarePatch}.

%%%%%%%%%%%%%%%%%%%%%%%%%%%%%%%%%%%%%%%%%%%%

%%%%%%%%%%%%%%%%%%%%%%%%%%%%%%%%%%%%%%%%%%

\section{Energy flux in tt-projection}
\label{sec:tt} 
In this section we compute the energy flux across hypersurfaces generated by the time-translation Killing vector $T$.  We exclusively work with tt-projection.
We start by observing some useful properties of the tt-projection,
\begin{eqnarray}
\partial_{\eta}(\chi_{ij}^\rom{tt}(\eta, r)) & = & (\partial_{\eta}\chi_{ij}(\eta, r))^\rom{tt},  \label{tt_commute1}
 \\
\partial_{r}(\chi_{ij}^\rom{tt}(\eta, r)) &=& (\partial_{r}\chi_{ij}(\eta, r))^\rom{tt}, \label{tt_commute2}
\eea
i.e., tt-projection commutes with $\partial_\eta$ and $\partial_r$. 
Moreover, 
\be
\partial_m(\chi_{ij}^\rom{tt}(\eta, r))  = 
(\partial_{m}\Lambda_{ij}^{~~kl})\chi_{kl}(\eta, r) +
\hat{x}_m(\partial_r\chi_{ij}(\eta, r))^\rom{tt} ,
\ee
as a result  we have, 
\bea 
\partial^j(\chi_{ij}^\rom{tt}(\eta, r)) & = &
\hat{x}^j\Lambda_{ij}^{~~kl}\partial_r\chi_{ij}(\eta, r) +
(\partial^{j}\Lambda_{ij}^{~~kl})\chi_{kl}(\eta, r) \nn \\ &=& \mathcal{O}(r^{-1}),
 \label{Transversality} 
 \eea
where we used,
\bea
\partial_{m}\Lambda_{ij}^{~~kl} & = & -
\frac{1}{r}\left[\hat{x}_i\Lambda_{mj}^{~~~kl} +
\hat{x}_j\Lambda_{mi}^{~~~kl} + \hat{x}^k\Lambda_{ijm}^{~~~~l} +
\hat{x}^l\Lambda_{ijm}^{~~~~k}\right] = \mathcal{O}(r^{-1}) \ . \label{ProjectorDerivative} 
\end{eqnarray}
The traceless-ness of $\chi_{ij}^\rom{tt}$ is manifest, but  
$\chi_{ij}^\rom{tt}$ satisfies the spatial transversality condition \eqref{TTconditions} to $\mathcal{O}(r^{-1})$ only, cf.~\eqref{Transversality}.

%%%%%%%%%%%%%%%%%%%%%%%%%%%%%%%%%%%%%%%%%%%%

\subsection{Energy flux across hypersurfaces of constant radial physical distance}

\label{energy_flux}

Hypersurfaces of constant radial physical distance can be defined as,
\be
 \Sigma_\rho: \qquad  \rho := a (\eta) r = - \frac{r}{H \eta} = \mbox{const}.
\ee
These hypersurfaces are generated 
by the time-translation 
Killing vector $T$, cf.~\eqref{time_translation}, \be T  = T^\a \partial_\a = - H (\eta \partial_\eta + r \partial_r).\ee
Let $\tau$ be the Killing parameter along the integral curves of the Killing vector $T$ satisfying,
\begin{align}
\frac{d\eta}{d\tau}&=-H\eta,&
 \frac{d x^i}{d\tau}&=-H x^i, & \label{KillingParameter}
\end{align}
then, coordinates on $\Sigma_\rho$ can be taken to be $\tau, \theta,$ and $\phi$.
The induced metric  on $\Sigma_\rho$ is, 
\be
h_{ab} = \rom{diag}(H^2\rho^2 - 1,  \rho^2, \rho^2  \sin^2\theta).
\ee 
This metric is of Lorentzian signature for $H\rho < 1$ (inside the cosmological horizon), is degenerate for $H\rho = 1$ (the
cosmological horizon),  and is of Euclidean signature for $H\rho > 1$ (outside the cosmological horizon); see figure~\ref{PoincarePatch}.  In this subsection we work with the timelike and spacelike cases; the case of  the null cosmological horizon is considered in the next subsection (section \ref{cosmological_horizon}).

  The volume element for the non-null cases is
\be
\sqrt{|h|} = \sqrt{|H^2\rho^2-1|} \ \rho^2 \sin\theta,
\ee
and the unit normal is
\be
n_{\a} = {\epsilon} \ a |
H^2\rho^2-1|^{-1/2}\left(H\rho, x_i/r\right).
\ee Here $\epsilon =
+1$ for time-like hypersurfaces $H\rho < 1$, and $-1$ for  space-like hypersurfaces $H\rho > 1$. 
Therefore, the infinitesimal volume element vector field is \cite{Poisson}
\be
d \Sigma_\a = \epsilon n_\a \sqrt{h} \ d^3 \xi  = a^{3} \, r^2 \, \sin\theta \ \left(H\rho, \frac{x_i}{r}\right) d \tau \, d\theta \, d \phi.
\ee
The hypersurface integral  \eqref{ET} for energy flux is then written as, 
\bea \label{flux_physical_radius}
E_T = -\int_{\Sigma_{\rho}}d \Sigma_{\alpha} 
j^{\alpha} & = & 
-\int_{-\infty}^{+\infty}d\tau\int_{S^2}d \Omega\ r^{2} a^{3} \bigg(H\rho  \ j^\eta + 
\frac{j^ i x_i}{r}\bigg),
\eea
where $\tau$ is the Killing parameter defined in \eqref{KillingParameter}.  Using \eqref{KillingParameter},
 this expression can be rewritten as,
 \be
E_T  =- \int_{\Sigma_
\rho}  a^{4} \left( j^\eta + \frac{1}{H \rho} j^r
\right) d^3 x. \label{omega_rho}
\ee
In this expression, both terms  diverge as $\eta 
 \rightarrow 0$, or as $\rho \to \infty$.  It is easily seen  from \eqref{SympCur1} and \eqref{SympCur2} that  $j^{\eta}$ term  diverges as $\eta^{-2}
 $ while $j^r$ term diverges as $\eta^{-1}$.  This situation is similar to $E_T$ evaluated on constant $\eta$ slices in \cite{ABKII}. We will see below that,  as in \cite{ABKII}, the divergent pieces turn out to be total derivative. 
 
 %%%%%%%%%%%%%%%%%%%%%%%%%%%%%%%%%%%%%%%%%%%%

\subsubsection*{$j^{\eta}$ contribution}

Let us first look at the $j^{\eta}$ part of integral \eqref{flux_physical_radius}, we call it $E_T^{(1)}$, 
  \bea
E_T^{(1)} &=&-\int_{-\infty}^{+\infty}d\tau\int_{S^2}d \Omega\ r^{2} a^{3} \left(H\rho  \ j^\eta \right) \\
&=&
 -\frac{1}{64\pi G} \int_{-\infty}^{+\infty} d \tau \int_{S^2} d \Omega \ r^{2} a^{3}  \ H\rho \
a^{-2}  \bigg[\partial_{\eta}\left[T \cdot \partial \chi_{ij}\right] \chi^{ij}-\partial_{\eta}\chi^{ij}
 (T \cdot \partial  \chi_{ij})\bigg]\\
 &=&\frac{1}{64\pi G} H^2 \rho^{3} \int_{-\infty}^{+\infty} d \tau \int_{S^2} d \Omega  \
 \Bigg[\eta \ \frac{d}{d\tau} \big[\partial_{\eta}\chi_{ij}\big] \chi^{ij}
 -H \eta \ \big [\partial_{\eta}\chi_{ij}\big] \chi^{}ij
 -\eta \ \partial_{\eta} \chi^{ij} \left[\frac{d} {d \tau} \chi_{ij}\right]\Bigg]\\
 %
% %
 &=&-\frac{1}{32\pi G}H^2 \rho^{3} \Bigg\{\int_{-\infty}^{+\infty} d \tau \int_{S^2} d \Omega  \
 \eta \, \left[ \partial_{\eta} \chi^{ij} \right] \left[ \frac{d} {d \tau} \chi_{ij} \right]
 -\frac{1}{2}  \int_{-\infty}^{+\infty} d \tau \int d \Omega  \
\frac{d}{d\tau} \bigg [\eta \ \big[\partial_{\eta}\chi_{ij}\big] \chi^{ij}\bigg]\Bigg\},\nn \\ \label{omega_eta_final}
  \eea
where we have done the following manipulations. In the first step we have 
 substituted \eqref{EFI}.  In the second  step we have used the property that 
 $T \cdot \partial = \frac{d}{d\tau}$ and $
\partial_{\eta}[\frac{d}{d\tau}]=\frac{d}{d\tau}[\partial_{\eta}]-H
\partial_{\eta}$. In the third step we 
have done integrations by part 
with respect to the Killing parameter $\tau$ and have made use of equation 
\eqref{KillingParameter}. This integration by parts is valid because $\frac{d}
{d\tau}$ is tangential to $\Sigma_\rho$. 

The second term in expression \eqref{omega_eta_final} is a total derivative. 
This integral is zero for the following reasons. On timelike $\rho$ = constant hypersurfaces, $
\tau=+\infty$ corresponds to  
 future timelike infinity $i^+ $ and $\tau=-\infty$ corresponds to  past timelike infinity $i^- $.  Assuming that the source is static at the boundary points \cite{ABKIII,DHI}, i.e., 
 $ \frac{dQ_{ij}}{d\tau}\big|_{\tau=\pm\infty}=0$ and $\frac{d\overline{Q}
 _{ij}}{d\tau}\big|_{\tau= \pm\infty}=0$, $\chi_{ij}$ vanishes at $i^+$ and $i^-$. Hence the end point contributions  in the integral \eqref{omega_eta_final} vanish for timelike hypersurfaces.
 
On spacelike $\rho$ = constant hypersurfaces, $\tau=+\infty$ corresponds to  
 future timelike infinity $i^+ $ and $\tau=-\infty$  corresponds to spatial infinity $i^0$;  see figure~\ref{PoincarePatch}. 
$\chi_{ij}$  vanishes at $i^0$ due to no incoming radiation boundary conditions at $\eta = - \infty$. Hence the end point contributions  in the integral \eqref{omega_eta_final} also vanish for spacelike hypersurfaces.

  %%%%%%%%%%%%%%%%%%%%%%%%%%%%%%%%%%%%%%%%%%%%

  \subsubsection*{$j^{i}$ contributions}
  
Let us now  look at the $j^{i}$ part of integral \eqref{flux_physical_radius}. We  call this piece $E_T^{(2)}$. Upon substituting \eqref{EFII} we get four terms. We separate the contributions of these terms based on their derivative structures. Two of these terms are,  $E_T^{(2, I)}$,
\bea
E_T^{(2, I)}&=&-\frac{1}{32\pi G}\int_{-\infty}^{+\infty} d\tau \int_{S^2} d \Omega \ r^{2} a^{3} a^{-2} \
\frac{x^k}{r} \bigg\{ \chi^{lm} \partial_{m}(T \cdot \partial) \chi_{kl}-
(T \cdot \partial) \chi^{lm} \partial_{m}\chi_{kl}
\bigg\}\\
&=& -\frac{\rho}{32\pi G}  \int_{-\infty}^{+\infty} d\tau \int_{S^2} d \Omega \
{x^k} \bigg\{ \chi^{lm} \partial_{m}(T \cdot \partial) \chi_{kl}-
(T \cdot \partial) \chi^{lm} \partial_{m}\chi_{kl}
\bigg\} \\
&=&-\frac{\rho}{32\pi G H} \int d^{3}x \ \frac{x^k}{r^3}
 \bigg\{ \chi^{lm} \partial_{m}(T \cdot \partial) \chi_{kl}-
(T \cdot \partial ) \chi^{lm} \partial_{m}\chi_{kl}
\bigg\}.
\eea  
Since we are working with tt-projection,  we have for the integrand,
 \begin{align}
 &{x^k}\left\{ \chi^{lm}_\rom{tt}~ \partial_m \left[ (T \cdot \partial) 
\chi_{kl}^\rom{tt} \right] - \left[ (T \cdot \partial) \chi^{lm}_\rom{tt} \right] \partial_m  \chi_{kl}
^\rom{tt} \right\}
\\
&= {x^k}\left\{ \chi^{lm}_\rom{tt}~ \partial_m \left[ (T \cdot 
\partial) (\Lambda^{ij}_{kl}~\chi_{ij}) \right]
-\left[ (T \cdot \partial) \chi^{lm}_\rom{tt} \  \partial_{m}(\Lambda^{ij}_{kl} \chi_{ij}) \right] \right\} \nn \\
&=r \ \hat{x}^k \left\{ \chi^{lm}_\rom{tt}\bigg[(\partial_{m}\Lambda^{ij}_{kl})
(T \cdot \partial) \chi_{ij}
+\Lambda^{ij}_{kl}~\partial_{m}(T \cdot \partial)\chi_{ij}\bigg]
-(T \cdot \partial) \chi_{lm}^\rom{tt} \bigg[ \Lambda^{ij}_{kl} \ (\partial_{m}\chi_{ij})
+\chi_{ij}(\partial_{m} \Lambda^{ij}_{kl})
\bigg] \right\} \nn \\
&= r \ \chi^{lm}_\rom{tt} \left\{-\frac{1}{r}\Lambda_{lm}^{ij} (T \cdot \partial)
\chi_{ij}\right\}- r \ (T \cdot \partial) \chi_{lm}^\rom{tt} \left\{ -\frac{1}{r}\Lambda^{ij}_{lm} \ 
\chi_{ij}\right\} \nn \\
&=- \chi^{lm}_\rom{tt} (T \cdot \partial)\chi_{lm}^\rom{tt}+\chi^{lm}_\rom{tt} (T 
\cdot \partial)\chi_{lm}^\rom{tt} \nn\\
&= 0,
 \end{align}
i.e., these two terms  cancel each other.

The remaining terms  $E_T^{(2,II)}$ in the $j^i$ integral are,
\bea
E_T^{(2,II)} &= & \frac{1}{64\pi G} \int_{-\infty}^{+\infty} d \tau \int_{S^2} d \Omega \ r^{2} a^{3} a^{-2} \ \frac{x^k}{r} 
 \bigg[\partial_{k}(T \cdot \partial) \chi_{ij}   \ \chi^{ij}-\partial_{k}\chi^{ij}
 (T \cdot \partial ) \chi_{ij}\bigg]\\
  &=& -\frac{1}{64\pi G} H\rho^2  \int_{-\infty}^{+\infty} d \tau \int_{S^2} d 
 \Omega\
 \bigg[\eta \ \frac{d}{d\tau} \left [\partial_{r}\chi_{ij}\right] \chi^{ij}
 -H\eta \left [\partial_{r}\chi_{ij}\right] \chi^{ij}
 -\eta \ \partial_{r} \chi^{ij} \left[\frac{d} {d \tau} \chi_{ij}\right]\bigg] \nn  \\ 
 &=& \frac{1}{32\pi G}H\rho^2 \Bigg\{ \int_{-\infty}^{+\infty} d \tau \int_{S^2} d \Omega \
 \eta \ \partial_{r} \chi^{ij} \left[\frac{d} {d \tau} \chi_{ij}\right]
 -\frac{1}{2}  \int_{-\infty}^{+\infty} d \tau \int_{S^2} d 
 \Omega  \
 \frac{d}{d\tau} \bigg[\eta \ \left [\partial_{r}\chi_{ij}\right] \chi^{ij}\bigg]
 \Bigg\},\nn \\ \label{omega_i_final}
\eea
where in arriving at these expressions we have done  manipulations similar to 
ones done above. In the first  step we have used the property that $T \cdot 
\partial = \frac{d}{d\tau}$ and $\partial_{r}[\frac{d}{d\tau}]=\frac{d}
{d\tau}[\partial_{r}]-H\partial_{r}$. In the second step we have done integrations by 
part with respect to the Killing parameter $\tau$ and have made use of 
equation \eqref{KillingParameter}.

The second term in expression \eqref{omega_i_final} is a total derivative. 
  On $\rho$ = constant surfaces, $\tau=+\infty$ corresponds to the 
 boundary point $i^+$. For $\rho$ = constant timelike 
 (spacelike) surfaces,  $\tau=-\infty$ corresponds to $i^-(i^0)$, see figure~
 \ref{PoincarePatch}. At all these points the field $\chi_{ij}$ vanishes. Hence contributions from 
 the total derivative term are zero in \eqref{omega_i_final}.

  \subsubsection*{Adding the two contributions}

The non zero contributions from $j^{\eta}$ and $j^{i}$ to the flux integral are
 \be
  E_T =  
  \frac{H\rho^{2}}{32\pi G}\left[\int_{-\infty}^{\infty} d\tau \int_{S^2} d\Omega \  
  \left[
  \frac{d}{d\tau} \chi_{ij}^\rom{tt}
  \right] 
  \left(
  r\partial_{\eta}\chi_{kl}^\rom{tt}+\eta \partial_{r}\chi_{kl}^\rom{tt}
  \right)
  \right] \delta^{ik}\delta^{jl}. \label{Flux_Phys}
  \ee
  At this stage we can use various 
  identities from section \ref{Identity} to get,
\bea
E_T &=&   \frac{G\rho^{2}}{8\pi} \int_{S^2} d\Omega \int_{-\infty}^{\infty} d\tau  \ \ \delta^{ik}\delta^{jl} \times \nn \\ 
& & \bigg[ \frac{R_{ij}^\rom{tt}R_{kl}^\rom{tt}}
  {\rho^2}+\frac{(1+H
  \rho)A_{ij}^\rom{tt}R_{kl}^\rom{tt}}{\rho^{3}}+\frac{H(1+H\rho)A_{ij}^\rom{tt}
  R_{kl}^\rom{tt}}{\rho^{2}}+\frac{H (1+H\rho)^2 
  A_{ij}^\rom{tt}A_{kl}^\rom{tt}}{\rho^{3}}\bigg], \label{energy_main_sec}
\eea
  where we recall that $A_{ij}$ is defined in \eqref{A_def} \be
  A_{ij}=\ddot{Q}_{ij}+2H\dot{Q}_{ij}+H\dot{\overline{Q}}_{ij}+2H^{2}
  \overline{Q}_{ij},\ee 
  and it satisfies identities \eqref{R_A_old}
  \be
  R_{ij}=\dot{A}_{ij}+HA_{ij}=\frac{d}{d
   \tau}A_{ij}-HA_{ij}.
\label{R_A}
   \ee
  In arriving at  expression \eqref{energy_main_sec}
   we have  used the fact that 
   $\partial_{r}$ and $\partial_{\eta}$ commute with the tt-projection, cf.~equations \eqref{tt_commute1}--\eqref{tt_commute2}. 
We also note that the operation of tt-projection
  commutes with the dot operation.
  
   Interestingly, all the other terms except the $RR$ term in expression  \eqref{energy_main_sec} combine into a  total derivative. Substituting $R_{ij}$ in terms of $A_{ij}$ in the other terms in \eqref{energy_main_sec} we get,
\bea
\bigg[\frac{(1+H\rho)^{2}}{\rho}  A_{kl}^\rom{tt}\bigg(\frac{d}{d\tau} A_{ij}
^\rom{tt}\bigg)\bigg]\delta^{ik}\delta^{jl} 
=\bigg[\frac{(1+H\rho)^{2}}{2\rho}\frac{d}{d\tau}\bigg( A_{ij}
^\rom{tt} A_{kl}^\rom{tt}\bigg)\bigg]\delta^{ik}\delta^{jl},
\eea
which is a total derivative on $\Sigma_\rho$. Like in the previous subsection, contributions from this total derivatives terms vanish. %see discussion below \eqref{total_derivative1}.
This is so because $A_{ij}$ vanishes due to the staticity assumption of the source at the boundary points.
Hence,  
these terms do not contribute to the energy flux. Note that, formally several 
of these total derivative terms do not have a good limit as $\rho \rightarrow 
\infty$, 
  reflecting the fact that the hypersurface integral of the symplectic current density itself does not have a good limit on $\mathcal{I}^{+}.$
However, the divergent terms turn out to be total derivatives, as in \cite{ABKII}. 
%%%%%%%%%%%%%%%%%%%%%%%%%%%%%%%%%%%%%%%%%%%%

A final expression is therefore,
\be
E_T =   \frac{G}{8\pi} \int_{S^2} d\Omega \int_{-\infty}^{\infty} d\tau  \ \left[ R_{ij}^\rom{tt}R_{kl}^\rom{tt} \right]\ \delta^{ik}\delta^{jl}. \label{energy_flux_final}
\ee

%%%%%%%%%%%%%%%%%%%%%%%%%%%%%%%%%%%%%%%%%%%%

%%%%%%%%%%%%%%%%%%%%%%%%%%%%%%%%%%%%%%%%%%%%

\subsection{Flux integral on cosmological horizon}

\label{cosmological_horizon}

The analog of the above computation can also be done on the cosmological horizon. 
The cosmological horizon is a null surface at,
\be
\cH^+ : \qquad \eta + r = 0.
\ee 
The fact that it is a null surface brings about some non-trivial changes to the computation of subsection \ref{energy_flux}, which we highlight below.
 On the cosmological horizon 
$\sqrt{h} =
H^{-2} \sin\theta,
$
and we fix the normalisation of the normal vector as, 
\be
n_{\mu} = -
|H\eta|^{-1}(1, x_i/r) ,
\ee 
so that $n^{\mu} = T^{\mu}$ at $\cH^+$. The flux integral is therefore,

 \bea
E_{T}=-\int_{\mathcal{H}^+} d\Sigma_{\a} j^{\a} &=&-\frac{1}{H^{3}} \int^{+\infty}_{-\infty} \frac{d\tau}{r} \int_{S^{2}} \bigg(j^{\eta}+\frac{j^i x_i} {r}\bigg). \label{energy_main_H}
\eea

\subsubsection*{$j^{\eta}$ contribution}

The $j^{\eta}$ terms in integral \eqref{energy_main_H} are
\bea
E_T^{(1)}&=&\frac{1}{64\pi G H}\int^{+\infty}_{-\infty}   d\tau 
\int_{S^2} d \Omega ~~ \eta~
\big(\chi^{ij} \partial_\eta \left[ (T \cdot \partial) \chi_{ij} 
\right]  - (T \cdot \partial) \chi_{ij}  \partial_\eta  \chi^{ij}\big).
\eea
Following the step similar to the previous subsection, 
this contribution becomes,
\bea
E_T^{(1)}&=&\frac{1}{32 \pi G H} \int^\infty_{-\infty} d \tau \int_{S^2} d \Omega \ r \left[ \left[\frac{d}{d\tau}\chi_{ij}^\rom{tt}\right] 
\ \partial_{\eta}\chi_{kl}^\rom{tt}\right] \delta^{ik} \delta^{jl}.
\eea

  \subsubsection*{$j^{i}$ contributions}
  
The $j^{i}$ part of integral again has two types of terms. The terms with the derivative structure of the form
\bea \label{tt_zero}
-\frac{1}{32 \pi G H}\int^{+\infty}_{-\infty}   d\tau \int_{S^2} d \Omega ~
x^k \left\{ \chi^{lm} \partial_m \left[ (T \cdot \partial) \chi_{kl} 
\right]   
- \left[ (T \cdot \partial) \chi^{lm} \right] \partial_m  \chi_{kl}  
 \right\} .
 \eea
 cancel with each other like in the  previous subsection.
The remaining terms in the integral become,
 \bea
E_T^{(2)} &=&-\frac{1}{64 \pi G H}\int_{-\infty}^{+\infty}   d\tau \int_{S^2} d \Omega 
\ \ \eta \left\{ \chi^{lm} \partial_r \left[ (T \cdot \partial) \chi_{lm} \right] 
- \left[ (T \cdot \partial) \chi^{lm} \right] \partial_r  \chi_{lm} \right\} \nn \\
&=& \frac{1}{32 \pi G H}\int   d\tau \int_{S^2} d \Omega 
 \ \ \eta \left[ (T \cdot \partial) \chi^{lm} \right] \partial_r \chi_{lm}\\
 &=&\frac{1}{32 \pi G H} \int d\tau \int_{S^2} d\Omega \ \eta \left[\ \left[\frac{d}
 {d\tau}\chi_{ij}\right]^\rom{tt} \partial_{r}\chi^{\rom{tt}}_{kl}\right]~\delta^{ik}\delta^{jl}.
\eea

   \subsubsection*{Adding the two contributions}

The energy flux across $\mathcal{H}^+$ is,
 \be
E_T =  \frac{1}{32 \pi G H}\int_{-\infty}^{\infty} d\tau \int_{S^2} d \Omega \left[\left[\frac{d}{d\tau}
 \chi_{ij} \right]^\rom{tt}(\eta \ \partial_{r}\chi_{kl}+r\partial_{\eta} \chi_{kl})^\rom{tt}
 \right] \delta^{ik}\delta^{jl}. \label{temp}
 \ee
 Now using the identities from section \ref{Identity} and substituting $H\rho=1$ on the 
 cosmological horizon, the  energy flux expression \eqref{temp} becomes,
  \begin{align} 
  E_T = \frac{G}{8\pi}\int_{S^2} d\Omega \int_{-\infty}^{\infty} d\tau \bigg[ R_{ij}^\rom{tt}R_{kl}^\rom{tt}
  +4H A_{ij}^\rom{tt}R_{kl}^\rom{tt}+4 H^{2}
  A_{ij}^\rom{tt}A_{kl}^\rom{tt}\bigg] \delta^{ik}\delta^{jl}.
  \end{align}
 Again terms other than the $RR$ term combine into a  total derivative. Using \eqref{R_A}, we note that,
 \be
4H  \int_{-\infty}^{\infty} d\tau \bigg[  A_{ij}^\rom{tt}R_{kl}^\rom{tt}+ H
  A_{ij}^\rom{tt}A_{kl}^\rom{tt}\bigg] \delta^{ik}\delta^{jl}
 = 2 H \int_{-\infty}^{\infty} d\tau \left[ \frac{d}{d\tau} \left[A_{ij}^\rom{tt}A_{kl}^\rom{tt}\right]\delta^{ik}\delta^{jl} \right] = 0.
 \ee
 Hence, the energy flux across $\mathcal{H}^+$ is simply,
 \be
 E_T 
 =  \frac{G}{8\pi} \int_{-\infty}^{\infty} d\tau \int_{S^2} d \Omega ~R_{ij}^\rom{tt}~R_{kl}^\rom{tt}~\delta^{ik}\delta^{jl}. \label{EF_CH}
 \ee

 \subsection{Sharp propagation of energy}
The integrands in integrals \eqref{energy_flux_final} and \eqref{EF_CH}  are exactly the same. In particular, the integrand is independent of $\rho$. Hence the power radiated
 \be
 P = \frac{dE}{d\tau} = \frac{G}{8\pi}  \int_{S^2}  d \Omega ~R_{ij}^\rom{tt}~R_{kl}^\rom{tt}~\delta^{ik}\delta^{jl} \label{power}
 \ee 
 is independent of $\rho$. 
 
 The power is a function of retarded time alone. Along the outgoing null rays, retarded time is constant, see figure~\ref{PoincarePatch}.  Specifically, the power can be computed at a cross-section of the cosmological horizon or at a cross-section of the future null infinity. As long as the cross-sections are on the same retarded time the two expressions are identical. This is the sense in which propagation of energy flux  is sharp in de Sitter spacetime.  See also \cite{DHII, Bonga} for related comments.

 \subsection{Comparison with the stress-tensor approach of \cite{DHII} }
 Reference \cite{DHII} also obtained an expression for the energy flux across hypersurfaces of constant radial physical distance. It uses the Isaacson stress-tensor approach. 
 To compare our energy flux expression \eqref{Flux_Phys} to theirs, we first use
 \be
 \frac{d}{d\tau} = (T \cdot \partial) = - H (\eta \partial_\eta +  r \partial_r),
 \ee 
 and then expand  out the resulting expression to get,
 \be
E_T = -\frac{1}{32\pi G} \int d\tau \int_{S^2} d\Omega  \ H^{2}\rho^{2}\eta r  \  \bigg\{ \partial_{\eta}
 \chi_{ij}\partial_{\eta}\chi_{kl}+\partial_{r}\chi_{ij}\partial_{r}\chi_{kl}
-\frac{1+H^2\rho^2}{H\rho} \  \partial_{r}\chi_{ij}\partial_{\eta}\chi_{kl}
 \bigg\} \delta^{ik}\delta^{jl}. \label{energy_flux_DH}
 \ee
 This expression matches with that of \cite{DHII}
 (equations (52) and (53)), modulo the `averaging'.   The averaging is part of the Isaacson stress-tensor approach.  

 Our analysis differs from \cite{DHII} in another technical aspect. In reference \cite{DHII}, to obtain energy flux in the form of equation \eqref{EF_CH} from 
 \eqref{energy_flux_DH}, 
 the approximation 
 \be
 \partial_{\eta}
 \chi_{ij} \simeq -\partial_{r}\chi_{ij} \label{approximation}
 \ee
  was used. From the computation of section \ref{energy_flux}, we note that this approximation is not needed. The terms it ignores combine into 
 a total derivative.

 \section{Discussion}
 
 \label{sec:disc}

We have explored propagation of energy flux in the future Poincar\'e patch of de Sitter spacetime. 
We computed energy flux integral on hypersurfaces of constant radial physical distance. We showed that in the tt-projection,  the integrand in the energy flux expression on the cosmological horizon is same as that on the other hypersurfaces of constant physical radial distance.
This strongly suggests that the energy flux propagates sharply in de Sitter spacetime.
 We also related our flux expression  to a previously obtained expression of \cite{DHII}, where a Isaacson stress-tensor approach was used.

 Our work can be extended in several directions. Perhaps the most pressing extension is to generalise our computations in TT-gauge and clarify their relation to \cite{ABKII, ABKIII}. %We have presented some preliminary work in this direction  in appendix \ref{app:TT}. 
 To systematically study this problem, it will be useful to carefully define the covariant phase space using hypersurfaces of constant radial physical distance. Such an approach offers advantages over \cite{ABKII, ABKIII}, as in this foliation, slices near future null infinity do not intersect source's worldvolume. Hence the covariant phase space based on homogeneous solutions of Einstein's equations is better defined. It can perhaps also be useful to compute the electric and magnetic parts of the Weyl tensors adapted to $\rho =$ constant slicing and write the flux expression in terms of these tensors. We hope to return to some of these problems in our future work.

 \subsection*{Acknowledgements} We thank  Ghanashyam Date and Alok Laddha  for discussions. We are grateful to Ghanashyam Date for carefully reading a version of the manuscript, and for his detailed comments. This research is supported in part by the DST Max-Planck partner group project ``Quantum Black Holes'' between CMI, Chennai and AEI, Golm.

  \appendix
 
\section{Addendum: Energy flux in TT gauge}
\label{app:TT}
In this appendix we evaluate expression \eqref{flux_physical_radius}  in TT-gauge. Although we are not able to match our final answer to that of \cite{ABKIII}, the computations involved are sufficiently interesting to include this discussion as an appendix. This appendix is not included in the journal version of the paper. 
For ease of reference we write the energy flux expression \eqref{flux_physical_radius} again,
%In section \ref{energy_flux} we saw that the energy flux across hypersurfaces of constant radial physical distance is written as, 
\bea 
E_T = -\int_{\Sigma_{\rho}}d \Sigma_{\alpha} 
j^{\alpha} & = & 
-\int_{-\infty}^{+\infty}d\tau\int_{S^2}d \Omega\ r^{2} a^{3} \bigg(H\rho  \ j^\eta + 
\frac{j^ i x_i}{r}\bigg), \label{flux_physical_radius_TT}
\eea
where recall that $\tau$ is the Killing parameter defined in \eqref{KillingParameter}. 

 %%%%%%%%%%%%%%%%%%%%%%%%%%%%%%%%%%%%%%%%%%%%

\subsubsection*{$j^{\eta}$ contribution}

Computation of the $j^{\eta}$ part of integral \eqref{flux_physical_radius_TT} is identical to the corresponding computation presented in section \ref{energy_flux}. A final answer is
 \bea
E_T^{(1)} 
 &=&-\frac{1}{32\pi G} \ H^2 \rho^{3} \int_{-\infty}^{+\infty} d \tau \int_{S^2} d \Omega  \
 \eta \, \left[ \partial_{\eta} \chi_{ij}^\rom{TT}  \right]\left[ \frac{d} {d \tau} \chi_{kl}^\rom{TT} \right] \ \delta^{ik} \delta^{jl}.
  \eea
   
  %%%%%%%%%%%%%%%%%%%%%%%%%%%%%%%%%%%%%%%%%%%%

  \subsubsection*{$j^{i}$ contributions}
  
Let us first look at the $j^{i}$ part of integral \eqref{flux_physical_radius_TT}. We  call this piece $E_T^{(2)}$. Upon substituting \eqref{EFII} we get four terms. We separate the contributions of these terms based on their derivative structures. Two of these terms are,  $E_T^{(2, I)}$,
\bea
E_T^{(2, I)}&=& \frac{1}{32\pi G}\int_{-\infty}^{+\infty} d\tau \int_{S^2} d \Omega \ r^{2} a^{3} a^{-2} \
\frac{x^k}{r} \bigg\{ \chi^{lm} \partial_{m}(T \cdot \partial) \chi_{kl}-
(T \cdot \partial) \chi^{lm} \partial_{m}\chi_{kl}
\bigg\}\\
&=& \frac{\rho}{32\pi G }  \int_{-\infty}^{+\infty} d\tau \int_{S^2} d \Omega \
{x^k} \bigg\{ \chi^{lm} \partial_{m}(T \cdot \partial) \chi_{kl}-
(T \cdot \partial) \chi^{lm} \partial_{m}\chi_{kl}
\bigg\} \\
&=&\frac{\rho}{32\pi G H} \int d^{3}x \ \frac{x^k}{r^3}
 \bigg\{ \chi^{lm} \partial_{m}(T \cdot \partial) \chi_{kl}-
(T \cdot \partial ) \chi^{lm} \partial_{m}\chi_{kl}
\bigg\}.
\eea  
Upon an integration by parts we get, 
\bea
E_T^{(2, I)}&=& \frac{1}{32 \pi G H} \bigg[ -\int d^3 x  \ \partial_{m} \bigg(\frac{\rho x^k}{r^{3}}\bigg) \
\chi^{lm}_\rom{TT} \  (T \cdot \partial) \chi_{kl}^\rom{TT}
+\int d^3 x  \ \partial_{m} \bigg(\frac{\rho x^k}{r^{3}}\bigg) \
\chi^\rom{TT}_{kl} \  (T \cdot \partial) \chi^{lm}_\rom{TT}
\bigg] \nn \\
&=&0,
\eea
i.e., these two terms exactly cancel each other in TT gauge since $\partial^{m} 
\chi_{lm}^\rom{TT}=0$.

For the remaining terms in the $j^i$ integral, computation is identical to the corresponding computation presented in section \ref{energy_flux}. A final answer is
\bea
E_T^{(2,II)} 
 &=& \frac{1}{32\pi G} \ H\rho^2 \int_{-\infty}^{+\infty} d \tau \int_{S^2} d \Omega \
 \eta \ \partial_{r} \chi_{ij}^{\rom{TT}} \left[\frac{d} {d \tau} \chi_{kl}^{\rom{TT}}\right] \delta^{ik} \delta^{jl}.
\eea

  %%%%%%%%%%%%%%%%%%%%%%%%%%%%%%%%%%%%%%%%%%%%

  \subsubsection*{Adding the two contributions}
A final expression for the energy flux in TT gauge is,
\bea
 E_T &=&  \frac{1}{32 \pi G}H\rho^{2}\left\{\int d\tau \int_{S^2} d\Omega \left[ \frac{d}{d\tau} \chi_{ij}
  ^\rom{TT}\right] 
  \big(r\partial_{\eta}\chi_{kl}^\rom{TT}+\eta \partial_{r}\chi_{kl}^\rom{TT}\big)\right\} \delta^{ik}\delta^{jl} \\
&=& \frac{G}{32 \pi} H\rho^{2}\left\{\int d\tau \int_{S^2} d\Omega \bigg[\frac{2}{\rho}R_{ij}+\frac{2H}{\rho}{(1+H\rho)} A_{ij}\bigg]^\rom{TT} 
  \bigg[ \frac{2}{H\rho}R_{kl}+\frac{2(1+H\rho)}{H\rho^{2}}A_{kl}
 \bigg]^\rom{TT}\right\}\delta^{ik}\delta^{jl}, \label{TT_energy}\nn \\
 \eea
where we have used the fact that $\partial_\eta$ and $r\partial_r$ commute with the TT operation.

Although we do not have a clear interpretation of \eqref{TT_energy}, neither a detailed understanding of its relation of \cite{ABKIII}, we make the following (possibly interesting/useful) observation. Under the integral sign, we can first evaluate the expressions at  $\rho$ = constant surface and then take its TT part\footnote{Although the TT conditions are tailored to $\eta =$ constant slices.}. Thought of it in this way,  it appears appropriate to pull out  factors of $\rho$ from bracketed expressions in \eqref{TT_energy}. Then, we can express energy flux as,
\be
E_T=  
  \frac{G}{8 \pi} \int d\tau \int_{S^2} d\Omega  \left[R_{ij}^\rom{TT}R_{kl}^\rom{TT}\right]\delta^{ik}\delta^{jl},
 \ee
as the remaining three terms in 
energy expressions can be written as a total derivative,
\bea
\frac{1}{2\rho} (1+H\rho)^{2} \frac{d}{d\tau}\bigg(A_{ij}^\rom{TT}A_{kl}
^\rom{TT}\bigg)
 \delta^{ik}\delta^{jl}.
\eea

\end{document}